\documentclass[12pt]{iopart}

\usepackage{graphicx}
\usepackage{mathrsfs}
\usepackage{arydshln}
\usepackage{accents}
\usepackage{color}

\usepackage{iopams}

\begin{document}

\title[Hierarchical structure of Hamiltonian systems]{Hierarchical structure of noncanonical Hamiltonian systems}

\author{Z Yoshida$^1$ and P J Morrison$^2$}
\address{$^1$  Graduate School of Frontier Sciences, University of Tokyo, Kashiwa, Chiba 277-8561, Japan}
\address{$^2$ Department of Physics and Institute for Fusion Studies, University of Texas at Austin, Austin, TX 78712, USA}
\date{\today}
\ead{yoshida@ppl.k.u-tokyo.ac.jp, morrison@physics.utexas.edu}

\begin{abstract}
Topological constraints play a key role in the self-organizing processes that create structures in macro systems.
In fact, if all possible degrees of freedom are actualized on equal footing without constraint, 
the state of `equipartition' may bear no specific structure.
Fluid turbulence is a typical example --  while turbulent mixing seems to increase entropy, a variety of sustained vortical structures  can emerge.
In Hamiltonian formalism, some topological constraints are represented by \emph{Casimir invariants} (for example, helicities of a fluid or a plasma),
and then, the effective phase space is reduced to the Casimir leaves.
However, a general constraint is not necessarily \emph{integrable}, which precludes the existence of an appropriate Casimir invariant; the circulation is an example of such an invariant.
In this work, we formulate a systematic method to embed a Hamiltonian system in an extended phase space; 
we introduce \emph{phantom fields} and extend the Poisson algebra.
A phantom field defines a new Casimir invariant, a \emph{cross helicity},
which represents a topological constraint that  is  not integrable in the original phase space.
Changing the perspective, a singularity of the extended system may be viewed as a subsystem 
on which the phantom fields (though they are \emph{actual fields}, when viewed from the extended system) vanish, i.e., the original system.
This hierarchical relation of degenerate Poisson manifolds enables us to see the
``interior'' of a singularity as a sub Poisson manifold.  
The theory can be applied to describe  bifurcations  and instabilities
in a wide class of general Hamiltonian systems
[Yoshida \& Morrison, Fluid Dyn. Res. \textbf{46} (2014), 031412].
\end{abstract}

\pacs{45.20.Jj, 47.10.Df, 02.40.Yy,52.35.We}

\maketitle



\section{Introduction}

Concepts of stability, transport, mixing, and turbulence are intimately related to invariant sets of governing  dynamical systems. In particular,  constants of motion foliate the phase space with invariant sets, reducing the dimension of a  system, and  in two-degree of freedom Hamiltonian systems even a single invariant set  can provide a barrier to transport as occurs for magnetic field lines of  confined plasma configurations  (e.g.\ \cite{Morrison2000}).  For the two-dimension Eulerian fluid  equations, an infinite-dimensional dynamical system,   the existence of the enstrophy is associated with the inverse cascade to large scale structures and explains features of turbulent spectra  
(e.g.\ \cite{KM1980}). Indeed the search for invariants and their consequences has  been the subject of research for literally centuries. 

Historically, the  use of Lyapunov functions expanded upon the work of Lagrange and Dirichlet for determining stability in Hamiltonian systems.  In the context of plasma and fluid dynamics many works appeared based on the  idea of using constants of motion to find sufficient conditions for stability.  For example,  Woltjer\,\cite{Woltjer58} used his helicity together with energy to determine stability in magnetohydrodynamics, while Oberman and 
 Kruskal\,\cite{KO58} determined stability using  a  generalized entropy and Gardner\,\cite{Gardner63} used his measure preserving rearrangement idea in the context of kinetic theory.  Later, these ideas were used by 
 Arnold\,\cite{arnold66} in fluid mechanics to determine stability of shear flows.  It wasn't until the 1980s that the invariants in a variety of stability arguments, which were  previously obtained by ad hoc means,  were seen to be a consequence of the degenerate Poisson brackets of the general Hamiltonian structure of continuum matter models in the Eulerian variable description (e.g.\ \cite{Morrison_AIP,Morrison1998}).  At this time constants of motion associated with Poisson bracket degeneracy were dubbed \emph{Casimir invariants}, and all helicity like variables in fluids and plasmas are of this kind.  Therefore, topological  of fluid mechanics and plasma physics are in reality a consequence of the particular noncanonical  Hamiltonian structure of models in these fields.  It is the study of such topological invariants and their association to noncanonical Poisson brackets that is the subject of this paper. 

The topological constraints of  helicity-like Casimir invariants  are often deemed to be the key element of  self-organizing mechanisms that create \emph{structures} in macro systems, the large scale vortices of the inverse cascade being an example.   In fact, if all possible degrees of freedom are actualized on equal footing without constraint,  one would expect the state of `equipartition' to occur and there to be no specific structure.
Fluid turbulence is a typical example --  while turbulent mixing seems to increase entropy, a variety of sustained vortical structures  can emerge.  By suppression of microscopic degrees of freedom, a macro system can behave very differently  from  the simple sum of microscopic elements  (or, the direct product of the microscopic phase spaces). In a more geometrical language,   macro system dynamics may be identified as lying in a submanifold (often called a \emph{leaf}) embedded in the total phase space spanning the microscopic degrees of freedom, where  the reduction of the dimension is due to topological constraints. 

If a topological constraint can be represented by a scalar invariant,
we may formulate a variational principle augmenting the energy to find some but not all equilibrium points\,\cite{Morrison1998} 
(see \cite{JBT} for examples pertinent to the magnetic helicity constraint)
or the entropy to find statistical equilibria\,\cite{YoshidaMahajan2014} (see \cite{RT1} for an example pertinent to adiabatic invariants) using Lagrange multipliers. We may also formulate a perturbation theory about  the topological constraint by including the scalar invariant into the Hamiltonian while  supplying a conjugate variable\,\cite{YM_BIRS}. As noted above, here we study topological constraints of Casimirs in the framework of Hamiltonian systems theory.   Because of these invariants the  state vector is constrained to move on a submanifold that is an intersection of the level-sets of  the Casimir invariants (which we call Casimir leaves).
Although Casimir invariants are first integrals of the Hamiltonian formulation,
they are not pertinent to symmetries of a specific Hamiltonian;
instead, Casimir invariants are  attributes of  the underlying Poisson bracket algebra, in particular, its  degeneracy. 

In general there are not enough Casimir invariants to obtain all equilibria from an augmented variational principle.  This is because the nullity of the noncanonical Poisson bracket that determines Casimir invariants  may have subtle singularities.    The ``integrability'' of the determining equation for  Casimir invariants is the main mathematical issue we explore here.   A general constraint is not necessarily {integrable}, which precludes the existence of an appropriate Casimir invariant; the circulation is an example of such an invariant\,\cite{YoshidaMorrison2014}.  In order to overcome the Casimir deficit, in this work we formulate a systematic method for embedding  a Hamiltonian system into  an extended phase space by introducing  \emph{phantom fields}  that  extend the Poisson algebra.   A special case of this  embedding was treated in our earlier work\,\cite{YoshidaMorrison2014} where the extension of the Poisson algebra was of the semi-direct product form; such fields were called mock fields, a special case of phantom fields.   A phantom field defines a new Casimir invariant, a \emph{cross helicity}, which represents a topological constraint that  was not integrable in the original phase space (see Morrison\,\cite{Morrison1987} for an early precursor to this idea).

We may reverse the  point of view, whereby   a  singularity in  the determining equation for  Casimir invariants gives rise to a \emph{singular Casimir invariant}\,\cite{YMD2013,Yoshida-IUTAM}.
Viewed  from the extended phase space, the original Poisson manifold is embedded as the  singularity of the extended system. One can probe the singularity (which contains an infinite number of degrees of freedom) with  the help of   singular Casimir invariants.
 
In this paper, we elucidate  the physical meaning of the singular Casimir invariants by invoking some explicit examples, while  the rigorous mathematical framework will be described elsewhere. Specifically, in section 
\ref{sec:HamRev} we review noncanonical Hamiltonian structure, describe the Casimir determining equations, and explain the Casimir deficit problem.  In section \ref{sec:phantom} we describe our embedding, giving rise to a hierarchy of vortex-like Poisson structures, in which we can explore topological constraints and submanifold singularity. Section \ref{sec:soliton} contains as an example the  nonlinear ion acoustic wave and associated soliton leaves.  Finally, we conclude in section \ref{sec:conclusion}.


\section{Noncanonical Hamiltonian systems (degenerate Poisson algebras)}
\label{sec:HamRev}

\subsection{Singularity of the Poisson operator}
A noncanonical Hamiltonian system may be written as
\begin{equation}
\frac{\rmd}{\rmd t} \bi{z} = \mathcal{J}(\bi{z}) \partial_{\bi{z}} H(\bi{z}),
\label{Hamilton_eq_2}
\end{equation}
where $\bi{z}$ is the state vector, a member of the phase space $X$ 
(here a Hilbert space endowed with an inner product $\langle ~,~ \rangle$;
in the later discussion, $X$ will be a function space, and then we will denote the state vector by $u$),
$H(\bi{z})$ is the Hamiltonian (here a real-valued functional on $X$), 
$\partial_{\bi{z}}$ is the gradient in $X$,
and $\mathcal{J}$ is the Poisson operator.
We allow $\mathcal{J}$ to be a function of $\bi{z}$ on $X$, and write it as $\mathcal{J}(\bi{z})$.
We assume that the bilinear product   
\[
\{ F,G \} =
\langle \partial_{\bi{z}} F(\bi{z}), \mathcal{J} \partial_{\bi{z}} G(\bi{z})\rangle
\]
is antisymmetric and satisfies the Jacobi identity;
then $\{~,~\}$ is a Poisson bracket.
By a \emph{noncanonical Hamiltonian system} we mean a Poisson algebra $C^\infty_{\{~,~\}}(X)$, together with a Hamiltonian function. 

A \emph{canonical} Hamiltonian system is endowed with a \emph{symplectic} Poisson operator where 
\[
\mathcal{J}_c = \left( \begin{array}{cc}
0 & I
\\
-I & 0
\end{array} \right).
\]
However, our interest is in noncanonical systems endowed with
Poisson operators  $\mathcal{J}$ that are inhomogeneous and degenerate
(i.e.,  $\textrm{Ker}(\mathcal{J}(\bi{z}))$ contains nonzero elements, and its dimension may change depending on the position in $X$).  Since $\mathcal{J}$ is antisymmetric, 
$\textrm{Ker}(\mathcal{J}(\bi{z}))=\textrm{Coker}(\mathcal{J}(\bi{z}))$,
and hence, every orbit is topologically constrained on the orthogonal complement of
$\textrm{Ker}(\mathcal{J}(\bi{z}))$. 

A functional $C(\bi{z})$ such that $\{C,G\}=0$ for all $G$ is called a Casimir invariant
(or an element of the \emph{center} of the Poisson algebra).
If $\mathrm{Ker}(\mathcal{J})=\{0\}$, the case for a canonical Hamiltonian system,  then there is only a trivial invariant $C=$ constant in the center.
Evidently, a  Casimir invariant $C(\bi{z})$ is a solution to the differential equation
\begin{equation}
\mathcal{J}(\bi{z}) \partial_{\bi{z}} C(\bi{z}) =0.
\label{Casmir-1}
\end{equation}
When the phase space $X$ has a finite dimension, say $n$, (\ref{Casmir-1}) is a first-order
partial differential equation.
If $\mathrm{Ker}(\mathcal{J}(\bi{z}))$ has a constant dimension $\nu$ in an open set $X_\nu\subseteq X$,
and $n-\nu$ is an even number,
we can integrate (\ref{Casmir-1}) in $X_\nu$ to obtain $\nu$ independent solutions,
i.e.,  $\mathrm{Ker}(\mathcal{J}(\bi{z}))$ is locally spanned by the gradients of 
$\nu$ Casimir invariants (Lie-Darboux theorem).
The intersection of all Casimir leaves (the level-sets of Casimir invariants)
is the effective phase space, on which $\mathcal{J}(\bi{z})$ reduces to a symplectic Poisson operator.

However, the general (global) integrability of (\ref{Casmir-1}) is a mathematical challenge.
In the next subsection, we see a simple example that epitomizes the problem.

\subsection{Example of singularity and non-integrable center}
\label{ex:singularity}
Let us consider a two-dimensional system $\bi{z}= ~^t(x,y)\in X=\mathbb{R}^2$ with
\[
\mathcal{J} = x \mathcal{J} _c.
\]
The plane $x=0$ is a singularity at which $\mathrm{Rank}(\mathcal{J} )$ drops by two. 
The kernel of $\mathcal{J}$ consists of two eigenvectors:
\begin{eqnarray}
& &\mathrm{Ker}(\mathcal{J} ) = \{ \alpha_x \bnu_x + \alpha_y \bnu_y;\, \alpha_x,\alpha_y\in\mathbb{R} \},
\nonumber \\
& & ~~~~~~~~ \left\{ \begin{array}{l}
\bnu_x=\delta(x)\bi{e}_x, 
\\
\bnu_y=\delta(x)\bi{e}_y.
\end{array} \right.
\label{kernel}
\end{eqnarray}
The first component $\bnu_x$ of the kernel is \emph{integrable} and produces a singular Casimir invariant:
\[
\bnu_x = \partial_{\bi{z}} C_{\mathrm{ex}}(\bi{z}),
\quad 
C_{\mathrm{ex}}(\bi{z}) = Y(x),
\]
where $Y(x)$ is the Heaviside step function (with the gap filled).
We call $C_{\mathrm{ex}}$ an \emph{exterior Casimir invariant}.
However, the second component $\bnu_y$ is not a closed 1-form, thus
it is not representable as a gradient of some scalar function, i.e., we cannot 
integrate $\bnu_y$ to produce a Casimir invariant.
Characterization of this peculiar  element of $\mathrm{Ker}(\mathcal{J} )$ is one of the main issues of this work.



\subsection{The Casimir deficit problem}
The problem and our strategy of study are summarized as follows:
\begin{enumerate}
\item
To see the essence of the problem, 
let us assume that $X$ is a cotangent bundle $T^*M$ of a finite-dimensional manifold $M$.
Suppose that $\bi{w}$ is a non-zero element of  $\mathrm{Ker}(\mathcal{J})$.
For $\bi{w}\in T^*M$ to be written as $\bi{w}=\partial_{\bi{z}} C ~(= \rmd C)$ with a Casimir invariant (0-form) $C$,
$\bi{w}$ must be an exact 1-from (or, for local integrability, it must be a closed 1-form).
This is a rather strong condition, for  one can easily construct a counter example that violates the integrability, e.g., 
 the example given in section\,\ref{ex:singularity}.
Our strategy for improving the integrability of (\ref{Casmir-1}) 
is to embed the Poisson manifold into  higher-dimensional spaces;
by adding extra components to $\bi{w}$, we may make it exact in a higher-dimension space.

\item
The point where the rank of $\mathcal{J}(\bi{z})$ changes is a singularity of (\ref{Casmir-1}),
from which singular (hyperfunction) solutions are generated\,\cite{Yoshida-IUTAM}.
However, the hyperfunction Casimir invariants fall short of spanning $\mathrm{Ker}(\mathcal{J})$; 
(see section\,\ref{ex:singularity}).
This   ``Casimir deficit'' problem suggests that the interior of the singularity
cannot be well described by hyperfunctions 
(which are, in fact, the cohomology class of the sheaf of holomorphic functions).
We may yet describe the interior of the singularity as a subsystem (or, a Poisson submanifold),
on which a non-integrable element of  $\mathrm{Ker}(\mathcal{J})$ may be integrable to define a
Casimir invariant of a ``reduced'' Poisson operator.

\item
Because models of fluids and plasmas are formulated on infinite-dimensional phase spaces, 
we have to develop an infinite-dimensional theory.
For these systems (\ref{Casmir-1}) is a functional differential equation,
and a singularity may cause an infinite-dimensional rank change.
The reader is referred to Yoshida, Morrison \& Dobarro\,\cite{YMD2013} for an example of a singular Casimir invariant generated by singularities in a function space.
\end{enumerate}


\section{Hierarchy of degenerate Poisson manifolds and phantom fields}
\label{sec:phantom}

Given our perspective of seeing  a singularity of a Poisson operator as a subsystem
in which a different dynamics may reside,   we may say that singularities define a hierarchy of Hamiltonian systems.
To this end we extend by introducing   {\em phantom fields} that  helps us to ``integrate'' the kernel of the Poisson operator
and connect the hierarchy of Hamiltonian systems.  To understand what is meant by this we start with an illustrative  example.

\subsection{Example: a hierarchy of vortex dynamics systems} 
\label{subsec:2D_example}

\begin{table}
\caption{Hierarchy of two-dimensional vortex  systems.
Here  ${[}a,b{]}= \partial_y a \partial_x b - \partial_x a \partial_y b$.}
\begin{center}
\begin{tabular}{c|c|c}
\hline
system & state vector & Poisson operator 
\\ \hline
(I)  & $\omega$          & $\mathcal{J}_{\mathrm{I}}= {[}\omega,\circ {]}$ 
\\ \hline
(II) & $ \left( \begin{array}{c} \omega \\ \psi \end{array} \right)$ & 
$\mathcal{J}_{\mathrm{II}}=\left( \begin{array}{cc} 
{[} \omega,\circ  {]} &  {[}  \psi,\circ {]}  \\
 {[}  \psi  ,\circ {]}  & 0     
\end{array} \right)$
\\ \hline
(III)& $ \left( \begin{array}{c} \omega \\ \psi \\ \check{\psi}\end{array} \right)$ & 
$\mathcal{J}_{\mathrm{III}}=\left( \begin{array}{ccc} 
 {[}  \omega,\circ {]}  &  {[}  \psi,\circ {]}  &  {[}  \check{\psi},\circ {]}  \\
 {[}  \psi  ,\circ {]}  & 0  & 0 \\
 {[}  \check{\psi}  ,\circ {]}  & 0  & 0
\end{array} \right)$ 
\\ \hline
\end{tabular}
\end{center}
\label{table}
\label{table:2D-hierarchy}
\end{table}

In Table\,\ref{table:2D-hierarchy} we compare the Hamiltonian formalisms of 
well-known examples of two-dimensional vortex dynamics systems\,\cite{Morrison_AIP,YoshidaMorrison2014,MH84,MM84}.
We denote by $\omega=-\Delta\varphi$ the vorticity with $\Delta$ being  the Laplacian 
and $\varphi\in H^1_0(\Omega)\cap H^2(\Omega)$ for the  two-dimensional Eulerian velocity field  
$\bi{V}=~^t(\partial_y\varphi,-\partial_x\varphi)$.
Given a Hamiltonian
\[
H_{{\rm E}}(\omega) = -\frac{1}{2} \int \omega \,  ( \Delta^{-1}\omega) \,\rmd^2 x,
\]
the system (I) is the vorticity equation for  Eulerian flow, 
\[
\partial_t \omega + \bi{V}\cdot\nabla\omega =0.
\]
The Casimir invariants of the system (I) are 
\[
C_0 = \int f(\omega)\,\rmd^2 x,
\]
where $f$ is an arbitrary $C^2$ function.

If $\psi$ is the Gauss potential of a magnetic field, i.e., 
$\bi{B}=~^t(\partial_y\psi,-\partial_x\psi)$,
and the Hamiltonian is
\[
H_{{\rm RMHD}}(\omega,\psi) = -\frac{1}{2} \int \left[\omega  \, ( \Delta^{-1}\omega) + \psi \, (\Delta\psi) \right]\,\rmd^2 x ,
\]
the system (II) is the reduced MHD system, 
\begin{eqnarray*}
& & \partial_t \omega + \bi{V}\cdot\nabla\omega = \bi{J}\times\bi{B},
\\
& & \partial_t \psi + \bi{V}\cdot\nabla\psi = 0. 
\end{eqnarray*}
In the system (II), $C_0$ is no longer a constant of motion, it being replaced by 
\begin{eqnarray*}
 C_1 &=& \int \omega g(\psi)\,\rmd^2 x, \\
 C_2 &=&  \int  f(\psi)\, \rmd^2 x.
\end{eqnarray*}

However, if the Hamiltonian $H$ is independent of $\psi$,
the dynamics of $\omega$ is unaffected by $\psi$,
while both $\omega$ and $\psi$ obey the same evolution equation.
Then, we call $\psi$ a \emph{phantom field} (previously called mock fields in \,\cite{YoshidaMorrison2014}), 
since it  can be chosen arbitrarily without changing the
dynamics of the \emph{actual field} $\omega$.
With  the special choice  $\psi=\omega$, both $C_1$ and $C_2$ evaluate as $C_0$,
i.e., $C_0$ is subsumed by $C_1$, the so-called cross helicity\,\cite{Fukumoto2008}, 
and $C_2$ as their special value
(indeed, $C_1$ and $C_2$ carry more information about system (I); 
in Sec.\,\ref{subsec:integrability}, we will see how the phantom field probes the topological constraints).
The constancy of $C_0$ is, then, due to the symmetry $\partial_\psi H =0$.
A modification of the Hamiltonian to involve $\psi$
destroys the constancy of $C_0$;
the electromagnetic interaction would be  a physical example of such a modification.
 
We can extend the phase space further to obtain a system (III) by adding another field $\check{\psi}$ 
that obeys the same evolution equation as $\psi$.
In the reduced MHD system, $\check{\psi}$ is a phantom field, i.e.,  it does not have a direct physical meaning;  
however, in the original RMHD context  such a field physically correspond to the pressure in the  high-beta MHD model\,\cite{MH84}.  
The Casimir invariants of this further extended system are
\begin{eqnarray*}
C_2 &= \int f(\psi)\,\rmd^2 x , \\
C_3 &= \int h(\psi\check{\psi})\,\rmd^2 x , \\
C_4 &= \int \check{f}(\check{\psi}) \,\rmd^2 x .
\end{eqnarray*}
For this system,  $C_1$ is replaced with the  new Casimir invariants $C_3$ and $C_4$.

\subsection{Integrability of topological constraints} 
\label{subsec:integrability}

An interesting consequence of extending the system from (I) to (II) is found in the
\emph{integrability} of the $\textrm{Ker}(\mathcal{J}_{\mathrm{I}})$.
In (I),
\[
\textrm{Ker}(\mathcal{J}_{\mathrm{I}}(\omega)) = \{ \psi;\,  [ \omega,\psi ] =0 \},
\]
which implies that $\psi$ and $\omega$ are related, invoking a certain scalar $\zeta(x,y)$, by
\begin{equation}
\psi = \eta(\zeta), \quad \omega= \xi(\zeta).
\label{general_Kernek}
\end{equation}
As far as $\xi$ is a monotonic function, we may write $\psi = \eta (\xi^{-1}(\omega))$, which
we can integrate to obtain the Casimir invariant $C_0(\omega)$ with $f(\omega)$ such that 
$f'(\omega) = \eta(\xi^{-1}(\omega))$.
Other elements of $\textrm{Ker}(\mathcal{J}_{\mathrm{I}}(\omega))$ 
that are given by nonmonotonic $\xi$ are not integrable and thus are unable to yield Casimir invariants.
Yet, we can integrate such elements as $C_1(\omega,\psi)$ 
in the extended  space of (II).  In fact, every member of $\textrm{Ker}(\mathcal{J}_{\mathrm{I}}(\omega))$ can be
represented as $\partial_\omega C_1=g(\psi)$ by choosing $\psi$ in $\textrm{Ker}(\mathcal{J}_{\mathrm{I}}(\omega))$.

Similarly, 
in the system (II),
we encounter the deficit of the Casimir invariant $C_2=\int f(\psi)\,\rmd^2x$
in covering all elements $\,^t(0,\chi)\in \textrm{Ker}(\mathcal{J}_{\mathrm{II}}(\omega,\psi))$ such that $ [ \psi,\chi ] =0$.
With  the help of a phantom field $\check{\psi}$, we can integrate every element of 
$\textrm{Ker}(\mathcal{J}_{\mathrm{II}}(\omega,\psi))$ as $C_3$.

\subsection{Submanifold of singularity}
In the preceding section, we extended the phase space to improve the integrability of the topological constraints (kernel elements).
Here we reverse the perspective, and see the ``singularity'' as a submanifold of a larger system.
We will introduce the  new notions  of \emph{exterior Casimir invariants} and \emph{interior Casimir invariants};
the former being a hyperfunction that identifies the singularity as a cohomology, 
while the latter turns out to be a Casimir invariant of the subsystem, which is invisible in the larger (extended) system.

Let us start with the system (II).
Apparently, the rank of $\mathcal{J}_{\mathrm{II}}$ drops (by infinite dimension) 
at the submanifold $\psi=0$.
This singularity is a leaf of the singular Casimir invariant (which we call an exterior Casimir invariant)
\[
C_{\mathrm{ex}} = Y(\|\psi\|^2),
\]
where $\|\psi\|^2 = \int |\psi|^2\,\rmd x^2$.
Notice that $C_{\mathrm{ex}}$ is a special singular form of the Casimir invariant $C_2(\psi)$.
This hyperfunction Casimir invariant has only one leaf $\psi=0$
(i.e., the equation $C_{\mathrm{ex}} =c$ has a solution, iff $c\in (0,1)$, and then the level-set is the singularity $\psi=0$), 
in marked contrast to other regular Casimir invariants
that densely foliate the phase space.

At the singularity $\psi=0$, $C_{\mathrm{in}} = \int f(\omega)\,\rmd x^2$ satisfies 
(denoting the state vector by $u=~^t(\omega,\psi)$)
\[
\left.
\mathcal{J}_{\mathrm{II}}\partial_u C_{\mathrm{in}} \right|_{\psi=0} = 0.
\]
Notice that  $C_{\mathrm{in}}$ is nothing but the Casimir invariant $C_0$ of the system (I), which, however, is not 
a Casimir invariant of the system (II).

From these observations, we draw the following conclusion:
a singularity of a Hamiltonian system (where $\mathrm{Rank}(\mathcal{J})$ drops) 
defines a submanifold, which can be a sub Hamiltonian system.\footnote{
In general, the singular submanifold may not bear a Hamiltonian structure.
For example, if the singularity is a one-dimensional manifold, it cannot  be a Hamiltonian system.}
Then, the Casimir invariants of the subsystem convert into the \emph{interior Casimir invariants}
of the larger system.
Simultaneously, the submanifold is identified as a singular leaf of the \emph{exterior Casimir invariant}
which is the hyperfunction of the cross helicity describing the coupling of the submanifold and the phantom fields.
The pair of the interior and exterior Casimir invariants constitute the \emph{singular Casimir invariants}.


\section{Another example: solitons on the vortex-free singular leaf}
\label{sec:soliton}
 
The ion acoustic wave is an example that may have a completely integrable structure described by the KdV (or KP) equation.
Here we show that the KdV (or KP) hierarchy is an internal structure on a singularity embedded in a general three-dimensional Hamiltonian parent model.  The symmetry is connected to an  internal Casimir element.

\subsection{Hamiltonian structure of ion acoustic waves}
\label{subsec:ion_acoustic_soliton}  
Let us start with the general three-dimensional formulation:
\begin{eqnarray}
& & \partial_t{\rho} = -\nabla\cdot(\bi{V} {\rho}),
\label{IAW-mass_conservation}
\\
& & \partial_t \bi{V}=-(\nabla\times\bi{V})\times\bi{V} - \nabla(\phi+V^2/2),
\label{IAW-momentum}
\\
& & -\Delta\phi = \rho - e^\phi ,
\label{IAW-Poisson}
\end{eqnarray}
where $\rho$ is the ion density, $\bi{V}$ is the fluid velocity, and
$\phi$ is the electric potential (which replaces the enthalpy of usual fluid models).
Instead of an equation of state, we use the nonlinear Poisson equation (\ref{IAW-Poisson}) to relate $\rho$ and $\phi$.
We formally solve (\ref{IAW-Poisson}) for $\phi$ as
\begin{equation}
\phi= \Phi(\rho).
\label{IAW-Poisson-sol}
\end{equation}

We may cast the system (\ref{IAW-mass_conservation})-(\ref{IAW-Poisson}) into Hamiltonian form\,\cite{Morrison_AIP}.
For the state vector $u=~^t(\rho,\bi{V})$,
we define
\begin{eqnarray}
H &=& \int_\Omega \left[ \frac{{\rho}V^2}{2} +  \frac{|\nabla\Phi|^2}{2} + (\Phi-1)\rme^\Phi \right]\,\rmd^3x ,
\label{IAW-Hamiltonian}
\\
\mathcal{J} &=& {\small \left( \begin{array}{cc}
0 & -\nabla\cdot 
\\
-\nabla & -{\rho}^{-1}(\nabla\times\bi{V})\times 
\end{array} \right) } . 
\label{IAW-J}
\end{eqnarray}
We assume that $u$ is a smooth function on $\Omega=\textrm{T}^3$ (i.e., periodic in every direction).
The phase space is $X=L^2(\Omega)$, and the domain on which the Poisson operator $\mathcal{J}$ is defined 
is a dense subset of $X$.
By the definition, we obtain
\[
\partial_u H = \left( \begin{array}{c}
V^2/2 + \phi
\\
\rho \bi{V}
\end{array} \right)\,;
\]
hence, the Hamiltonian system $\partial_t u = \mathcal{J} \partial_u H$ reproduces  (\ref{IAW-mass_conservation})-(\ref{IAW-Poisson}).

The Poisson operator (\ref{IAW-J}) has Casimir elements
\begin{eqnarray}
C_1 &=& \int_\Omega  {\rho}\,\rmd^3x ,
\label{IAW-Casimir-mass}
\\
C_2 &=& \frac{1}{2} \int_\Omega  (\nabla\times\bi{V})\cdot\bi{V} \,\rmd^3x ,
\label{IAW-Casimir-helicity} 
\end{eqnarray}
which, respectively, measure the total particle number and the helicity.

\subsection{Structures on a leaf of a singular Casimir invariant} 
\label{subsec:soliton}
When $\nabla\times\bi{V}\equiv 0$ (i.e., $\bi{V}$ is irrotational everywhere in $\Omega$)\footnote{
Somewhat subtler singularity occurs when $\nabla\times\bi{V}$ vanishes locally in $\Omega$;
see\,\cite{YMD2013}.
}, the rank of $\mathcal{J}$ drops (by an infinite dimension),
thus, the submanifold $\Gamma \subset X$ on which $\|\nabla\times\bi{V}\|=0$ is a singularity of $\mathcal{J}$.
By Kelvin's circulation theorem, an irrotational flow remains irrotational for all time,
i.e., an irrotational flow resides on the leaf $\Gamma$.
The determining Casimir invariant of $\Gamma$ is a singular functional
\begin{equation}
C_{\mathrm{ex}} = Y(\|\nabla\times\bi{V}\|^2) .
\label{IAW-Casinir-ex}
\end{equation}
The leaf of $C_{\mathrm{ex}}$ exists only at the singularity $\| \nabla\times\bi{V}\|=0$, 
unlike those of regular Casimir elements which fill the phase space.
Such a leaf $\Gamma$ is a submanifold (in fact, a linear subspace) included in the leaf of $C_2=0$
(the zero helicity may occur even when the vorticity is finite,
thus $\Gamma$ is a subset of the leaf of $C_2=0$).

On the singular Casimir leaf $\Gamma$ (the set of points satisfying $C_{\mathrm{ex}} \in [0,1)$), remarkable structures emerge,
which are characterized by an interior Casimir element:
\begin{equation}
C_{\mathrm{in}} = \int_\Omega  \bi{B}\cdot\bi{V}\,\rmd^3x ,
\label{IAW-Casinir-in}
\end{equation}
where $\bi{B}$ is an arbitrary divergence-free vector.
The simplest choice is a fixed unit vector.
With $\bi{B}=\bi{e}_x$, $C_{\mathrm{in}} = \int_\Omega  {V}_x\,\rmd^3x$ 
is the momentum in the direction of $\bi{e}_x$,
which plays an important role in characterizing the ion acoustic soliton propagating along the $x$-axis.

Let us delineate how the \emph{integrable structure} of ion acoustic waves is 
embedded  in the phase space $X$ of  general three-dimensional vorticical dynamics.
We consider a case where the fields included in $H$ are spatially one-dimensional, 
i.e., $\bi{V}=V(x,t)\bi{e}_x$ and $\rho=\rho(x,t)$.
Evidently, $\nabla\times\bi{V}\equiv 0$, thus such orbits are constrained on $\Gamma$.
Upon setting $\nabla\times\bi{V}\equiv 0$ this bracket becomes
\[
\{F,G\}=-\int  \left(
F_{\rho} \nabla\cdot G_{\bi{V}}- G_{\rho} \nabla\cdot F_{\bi{V}}
\right)\,\rmd^3x .
\]
Reducing to 1D it becomes
\[
\{F,G\}=-\int \left(
F_{\rho} \partial_x G_V- G_{\rho} \partial_x  F_{V}
\right)\,\rmd x .
\]
Finally, upon setting $\rho- \rho_0= V= w$  gives 
\[
\{F,G\}_{G} =-\int \left(
F_{w} \partial_x G_w- G_{w} \partial_x  F_{w}
\right)\,\rmd x 
= 2 \int ( G_{w} \partial_x  F_{w} )\,\rmd x ,
\]
which is the Gardner bracket for KdV.
Evidently, $\int {w}\,\rmd^3x$ is a Casimir invariant of $\{~,~\}_G$,
which is nothing but the interior Casimir invariant $C_{\mathrm{in}} $ with $\bi{B}=\bi{e}_x$,
and is  the first invariant of the KdV hierarchy.


\subsection{Introduction of a phantom and observation from an extended phase space}
As a more nontrivial extension of $\bi{B}$ defining the interior Casimir element (\ref{IAW-Casinir-in}),
we may regard $\bi{B}$ as a phantom field (2-form) co-moving with the fluid, i.e., a vorticity-like field 
obeying
\begin{equation}
\partial_t\bi{B} = \nabla\times (\bi{V}\times \bi{B}).
\label{induction}
\end{equation}
By extending $\mathcal{J}$ as
\begin{equation}
\tilde{\mathcal{J}} = {\small \left( \begin{array}{ccc}
0 & -\nabla\cdot & 0
\\
-\nabla & -{\rho}^{-1}(\nabla\times\bi{V})\times & -{\rho}^{-1}(\nabla\times\circ)\times\bi{B}
\\
0 & \nabla\times (\circ\times \rho^{-1}\bi{B}) & 0
\end{array} \right) } ,
\label{IAW-J-extended}
\end{equation}
we may consider an extended dynamics of $\tilde{u}=~^t(\rho,\bi{V},\bi{B}) \in \tilde{X}$;
as far as $\bi{B}$ is not included in the Hamiltonian, the dynamics of $u=~^t(\rho,\bi{V})$ is unaffected by the phantom field $\bi{B}$.

We find that (\ref{IAW-J-extended}) is the Poisson operator of magnetohydrodynamics
when $\bi{B}$ is the magnetic field\,\cite{MG80,YoshidaMorrison2014}.
The Casimir invariants are the total mass
$C_1 = \int {\rho}\,\rmd^3x $,
the cross helicity $C_X = \int  \bi{B}\cdot\bi{V}\,\rmd^3x$,
and the magnetic helicity
$C_M = \int \bi{B}\cdot\bi{A}\,\rmd^3x$ (where $\bi{A}$ is the vector potential; $\nabla\times\bi{A}=\bi{B}$).
On the singular leaf $\Gamma$, the cross helicity $C_X$ reads as the interior Casimir invariant (\ref{IAW-Casinir-in}).
Notice that $\bnu = ~^t(0,\bi{B}) \in \mathrm{Ker}(\mathcal{J}|_\Gamma)$ ($\nabla\cdot\bi{B}=0$) is 
not integrable in $X$ (like $\bnu_y$ of the example given in subsection\,\ref{ex:singularity}),
but it can be integrated as the cross helicity $C_X$ in the extended phase space $\tilde{X}$
(as well as $C_{\mathrm{in}}$ on $\Gamma$).


\section{Conclusion}
\label{sec:conclusion}

By embedding a Poisson manifold of a noncanonical Hamiltonian system into a higher-dimensional phase space, 
we showed how to  delineate topological structures within  a simple  picture.
In section \ref{sec:phantom}, 
we invoked   two-dimensional vortex systems to
explain the systematic method for constructing a hierarchy of Poisson manifolds;
we introduced \emph{phantom fields} and extended the Poisson algebra so that the
phantom fields are Lie-dragged by the flow vector. 
A phantom field was seen to define  a new Casimir invariant, a \emph{cross helicity},
which represents topological constraints that are not integrable in the original smaller phase space;
the \emph{circulation} being  such an example\,\cite{YoshidaMorrison2014}.

Representing a topological constraint by a scalar (Casimir invariant) brings about a number of advantages,
because we may deal with it in a  Hamiltonian (or  action) formulation. 
Multiplying a Casimir invariant by a constant (Lagrange multiplier) and adding it to the Hamiltonian,
we may define an \emph{energy-Casimir function} and apply it to the study of stability as a Lyapunov function.
Or, we can view a Casimir invariant as a momentum (or an action variable) 
and supply a conjugate coordinate (or an angle) variable to make a canonical pair.
By including the new conjugate variable into the Hamiltonian, we can ``unfreeze'' the   Casimir invariant,
and remove the topological constraint.
For example, the tearing-mode instability\,\cite{FKR63,White83} emerges when the topological constraint on the helical magnetic flux
(or the circulation of the magnetic field along a ``resonant'' helical loop) is removed by a finite parallel electric field\,\cite{YM_BIRS,YD2012}.
 

Alternatively, we saw that a smaller system can be viewed as a singular Casimir leaf of a larger system, 
such a leaf being  a singularity of the Poisson operator at which the rank changes.
The corresponding Casimir invariant was shown to be a hyper function generated by the singularity.
In section\,\ref{sec:soliton}, we described  the example of ion acoustic waves, where the singular Casimir leaf 
is the subspace of irrotational flows in which the solitons reside.
The theory of singular Casimir invariants and phantom fields
can be applied to a very large class of fluid and plasma dynamics (cf.\,\cite{YoshidaMorrison2014}), and in future publications we will explore other examples.

\ack
One of the authors (ZY) acknowledges stimulating discussions with Dr. Hosam Abd El Razek during his visit to Tokyo.
The work of ZY was supported by JSPS KAKENHI Grant Number 23224014 and 15K13532, while the 
work of PJM was supported by the U.S. Department of Energy Contract DE-FG02-04ER54742.

\end{document}